\documentstyle[preprint,aps]      {revtex}

\begin{document}
\setcounter{page}{1}
\pagestyle{plain}
\noindent
\centerline{\bf\Large Quantum Hall Fluid of Vortices in a Two Dimensional}
\centerline{\bf \Large Array of Josephson Junctions}
\\ \\
\noindent
\centerline{Ady Stern}
\centerline{Physics Department, Harvard University, Cambridge, Ma. 02138}
\noindent
\\ \\
\centerline{\large\bf Abstract}

A  two
dimensional  array of Josephson junctions in a magnetic field
is considered. It is shown that the dynamics of the vortices in
the array resembles that of electrons on a
two--dimensional lattice put in a magnetic field perpendicular
to the lattice. Under appropriate conditions,
this resemblance results in the formation of a quantum Hall fluid of
vortices. The bosonic nature of vortices and their long range
logarithmic interaction make some of the properties of the
vortices' quantum Hall fluid different from those of the electronic one.
Some of these differences are studied in detail. Finally, it is shown that
a quantum Hall fluid of vortices manifests itself in a
quantized Hall electronic transport in the array.

\noindent
\\ \\

\noindent
{\large\bf 1. Introduction}

This paper discusses  a quantized Hall effect (QHE)
state of vortices  in a two dimensional (2D) array of
Josephson junctions. Motivated by the analogy between Magnus force
acting on a vortex moving in a two--dimensional ideal fluid and Lorenz
force acting on a charge in a magnetic field,
we study the transport of
vortices in a Josephson junction array. In particular, we focus on the case
 in which the charging energy of the array is
minimized when the number of Cooper pairs on each element of
the array
is not an  integer.
We find that for a certain range of parameters the vortices are expected
to form a quantum Hall fluid, and the resistivity of the array
is expected to show a QHE behavior. While some of the properties
of the  quantum Hall fluid formed by the vortices are similar to those of
the well known Laughlin fluid, formed by electrons in QHE systems, we
find that the logarithmic interaction between the vortices leads to
interesting modifications of other properties.

The paper is
organized in the following way: in Section (2)
 we review the classical and quantum mechanical
analogies between the 2D
dynamics of
charged particles under the effect of a magnetic field and that of
vortices in a
2D fluid. These analogies, arising from the
analogy between Magnus and Lorenz forces,  motivate the introduction of
the system we analyze -- a  Josephson junction array in a magnetic
field, and the study of a quantized Hall effect in that system.
Section (2) is concluded with a precise formulation of the problem
to be studied.
In Section (3) we analyze the transport of vortices in this array.
We show that the dynamics of the vortices can be mapped on that of charged
particles under the effect of a magnetic field, a lattice--induced
periodic potential and a mutual interaction. The mutual interaction is
composed of a  logarithmic "static" part  as well as a
velocity--dependent  short ranged part. In Section (4) we analyze
the quantized Hall effect associated with the transport of the vortices, and
 its observable consequences. In particular, we study the
unique features of the QHE for logarithmically interacting particles.
Conclusions are presented in Section (5).

The possibility of  Quantum Hall phenomena in Josephson junction
arrays was recently discussed in two other works, one of Odintsov and
Nazarov \cite{Odintsov},
and the other of Choi\cite{Choi}. The regime of parameters
we consider is different
from the ones considered by these authors. We comment briefly on this
difference and its implications in section (2).

\vskip 1cm \noindent
{\large\bf 2.  Transport of vortices in a Josephson junction array --
introduction and motivation}

The classical dynamics of vortices in two--dimensional ideal fluids
is well known to resemble that of charged particles
under the effect of a strong  magnetic
field \cite{Lamb}.
An electron in a magnetic field is subject to Lorenz force, while
a vortex in an ideal fluid is subject to  Magnus force. Both forces are
proportional and perpendicular to the velocity. Two electrons in a strong
magnetic field  encircle each other, and so do two vortices in an
ideal fluid. The dynamics of both are well approximated by an
Hamiltonian that includes only a potential energy $V(x,y)$, where $x,y$ are
the planar coordinates, and for which  $x$ and $y$ are canonically
conjugate. This approximation is known as the guiding center approximation
for the electronic problem, and as Eulerian dynamics for the vortex
problem. This close resemblance  naturally raises the possibility of
 analogies between transport phenomena of electrons in a magnetic
field and those of  vortices in ideal fluids.

A vortex in a fluid can be viewed as an excitation in which each
fluid particle is given an angular momentum $l$ relative to the vortex
center \cite{Feynman}.
Consequently, the velocity field ${\vec v({\vec r})}$ of the fluid
satisfies $\int_\Gamma {\vec v}\cdot d{\vec l}={{2\pi l}\over m}$, where
$m$ is the mass of a fluid particle, and $\Gamma$ is a curve that
encloses the vortex center.  When the  vortex center moves with a velocity
$\vec u$, and the fluid is at rest far away from the vortex center,
the vortex center  is subject to a Magnus force, given by
$F_{Magnus}=2\pi l n {\vec u}\times{\hat z}$, where
$n$ is the number density of the fluid far away from the vortex core.
 Being both proportional and perpendicular
to the velocity of the vortex center, Magnus force obviously resembles
the Lorenz force acting on an electron moving in the $x-y$ plane under the
effect of a magnetic field $B\hat z$. This Lorenz force is given by
$F_{Lorenz}={{eB}\over c}{\vec u}\times {\hat z}$, where $\vec u$ is
the velocity of the electron.
Thus, the role played by the product ${e\over c}B$
 in the latter case is played by the product $2\pi l n$
 in the former. While the fluid density plays a role analogous to that
of a magnetic field, a fluid current plays a role analogous to that of an
electric field. To see that,  note that in
 a frame of reference in which the electron is at rest,
the force it is subject to looks as if it
 arises from an electric field, given by ${B\over c}{\vec u}\times{\hat z}$.
 Similarly, in
a frame of reference in which the vortex center is at rest, the force it is
subject to seems to arise from the motion of the fluid. Since the fluid
current density is  $\vec J=n\vec u$, in the vortex rest frame
the Magnus force is $F_{Magnus}=2\pi l
{\vec J}\times{\hat z}$, and ${\vec J}\times{\hat z}$ plays a role
analogous to that
of an electric field.
Thus, while the
fluid density affects the vortex dynamics in the same way a
magnetic field affects electronic dynamics, the fluid current plays
the role of an electric field.
Maxwell's equation ${\vec\nabla}\times{\vec E}
+{{\partial B}\over{\partial t}}=0$ is  then analogous to the continuity
equation in the fluid ${\vec\nabla}\cdot{\vec J}+
{{\partial n}\over{\partial t}}=0$. \cite{Thouless}\cite{Orlando}

 Quantum mechanics introduces two new ingredients to the analogies
 discussed above. The first is the quantum  of angular momentum $l$, given
 by $\hbar$ (or alternatively, the quantum of vorticity,
$h\over m$) \cite{Onsager}.
 The second is the
quantization of the magnetic flux,
 the integral of the magnetic field over area.
 This quantization is most clearly seen
through the Aharonov--Bohm effect \cite{Aharonov}:
 the Aharonov--Bohm phase shift
accumulated by an electron
 traversing a closed path in a magnetic field is
$2\pi$ times the number of flux quanta it encircles. Combining these
two ingredients together,
one should expect a quantization associated with
the integral of the number density over area, i.e.,
 with the number of particles. This
quantization should manifest itself in the phase accumulated by a
vortex carrying a single quantum of angular momentum, $\hbar$, when it
traverses a closed path in a fluid. Indeed, as shown first by Arovas,
Schrieffer and Wilczek \cite{Arovas},
  such a vortex does accumulate a geometric (Berry) phase \cite{Berry},
 and this
phase is
 $2\pi$ times the number of fluid
particles it encircles. {\it The analog of a flux quatum is then  a
single fluid particle}\cite{MPAFisher}.
Note that the analogy between vortex dynamics in
a fluid and electron dynamics in a magnetic field does not depend on the
fluid being charged, and is valid for neutral fluids as well.

Quantum transport of 2D electrons in a magnetic field crucially depends
on the
electronic filling factor, the ratio between the density of conduction
electrons and the density of flux quanta. For a very low filling factor,
$(\ll 1)$,
electrons are expected to form a Wigner lattice. At higher filling factors,
the quantized Hall effect takes place \cite{Prange}. Similarly, we expect
transport
of vortices in a fluid to depend on a vortex "filling factor",
the ratio between the density of
vortices and the density of fluid particles. However, in continious
two dimensional fluids this ratio is usually much smaller than one, and
the vortices indeed form an Abrikosov lattice.

How can the vortices "filling factor" be made larger? In this work we
make the vortex filling factor larger by considering a lattice
structure. As is well known, all properties of electrons on a lattice
are invariant to the addition of a magnetic flux quantum to a lattice
plaquette. Similarly, when we consider vortex
transport on a lattice-structured fluid, we find all properties to
be invariant to the addition of a single fluid particle to a lattice site.
It is this periodicity that allows us to make the effective filling factor
much larger than the ratio between the density of vortices and the density
of fluid particles.

Based on the foregoing general considerations, we study in this paper
a Josephson junction array
in a magnetic field. Josephson junction arrays were
extensively studied in recent years \cite{Mooij}\cite{Eckern}.
The array we consider is composed
of   identical small super--conducting dots coupled by a
nearest--neighbors Josephson coupling $E_J$, and by a capacitance matrix
$\hat C$. For definitness, we consider a square lattice of the
superconducting dots. Generalizations to other lattices are straight
forward.
A perpendicular magnetic field  induces  vortices
in the configuration of the superconducting phase. We denote the average
number
of vortices per lattice plaquette by ${\overline n}_v$.
Each of the dots carries a dynamical number of Cooper pairs, denoted by
$n_i$ (for the $i$'th dot), as well as positive background charges.
The charging energy of the array is minimized for a certain set of
values of $n_i$, which we denote by $n_{x,i}$. (In our notation, charge
is always expressed in units of $2e$, i.e., $n_i,n_{x,i}$ are
dimensionless.) While the $n_i$'s  are operators
with integer eigenvalues, $n_{x,i}$ are  real parameters, that are closely
related to the chemical potential of the dots.   In this work we consider
the case in which for all sites $n_{x,i}=n_x$.
The Hamiltonian describing the array is \cite{Eckern},
\begin{equation}
H={{(2e)^2}\over{2}}{\sum_{ ij}} (n_i-n_x){\hat C}^{-1}_{ij}(n_j-n_x)+
E_J{\sum_{\langle ij\rangle}}
(1-\cos(\phi_i-\phi_j-\int_i^j {\vec A}\cdot {\vec dl}))
\label{ham}
\end{equation}
where ${\sum_{\langle ij\rangle}}$ denotes a sum over nearest neighbors,
$n_i$ is the number of
Cooper pairs on the $i'$th dot, $\phi_i$ is the phase of the
superconducting order parameter on the $i$'th dot, $\vec A$ is the
externally put vector potential and the integral is taken between the
sites $i$ and $j$. A factor of ${2e}\over c$ is understood to be absorbed in
$\vec A$. The matrix ${\hat C}^{-1}$ is the
inverse of the capacitance matrix $\hat C$.
Generally, the matrix $\hat C$ includes  elements
coupling a dot to its nearest neighbors, to the substrate and
to  neighbors further away. The matrix elements of both $\hat C$
and ${\hat C}^{-1}$  are a
function of the distance between the sites $i$ and $j$. For short
distances the electrostatic energy is determined by nearest neighbors
capacitance only, and all other capacitances can be ignored.
The $ij$ matrix element of $C^{-1}$ is then ${{2\pi}\over{C_{nn}}}\log
{|r_i-r_j|}$, where $C_{nn}$ is the nearest neighbors capacitance
\cite{Eckern}.
For large distances, the electrostatic interaction depends also on
capacitance to the substrate and capacitance to neighbors further
away. The inverse capacitance matrix then decays with the distance.
Throughout
most of our discussion we assume that the size of the array is small
enough such that the charging energy is determined by nearest neighbors
capacitance only.  Then, the charging energy involves one energy scale,
$E_C\equiv{{e^2}\over {2C_{nn}}}$.
The effect of other capacitances is briefly discussed in
section (3).

Since our main interest in this study is focused on transport phenomena
of vortices, we constrain ourselves to arrays in which $E_J\widetilde{>}
E_C$. In that regime of parameters vortices are
mobile enough not to be trapped within plaquettes, but their rest energy
is large enough such that quantum fluctuations of vortex--antivortex
pair production can be neglected. Arrays in which $E_J\widetilde{>}
E_C$ were studied experimentally by van der Zant {\it et.al.}
\cite{Zant}, and were found to show a magnetic field tuned transition
from an almost super conducting state to an almost insulating state.
At weak magnetic fields the density of vortices is low, and their ground
state is the Abrikosov lattice. The array is then super--conducting.
The transition to the insulating state, at a critical value of the
magnetic field, is interpreted as caused by a transition of the
vortices from a lattice phase to a correlated super--fluid--like phase
\cite{MPAFisher}\cite{Zant}.

As mentioned in section (1), QHE phenomena in Josephson junction arrays
were discussed in two recent preprints. The first, by Odintsov and
Nazarov \cite{Odintsov},
focuses on the regime $E_C\gg E_J$, and discusses a quantum Hall
fluid of Cooper--pairs. The second, by Choi \cite{Choi},
focuses on the regime
$E_J\gg E_C$, and discusses a quantum Hall fluid of vortices. The quantum
fluid we discuss in this paper has some similarity to the one discussed
by Choi. However, the difference in the regime discussed, as well as our
detailed study of the effect of the logarithmic vortex--vortex interaction,
make some of our conclusions different from those of Choi.

 Due to the lattice structure of the array, the spectrum and eigenstates
of the Hamiltonian (\ref{ham}) are manifestly periodic with respect to
$n_x$, with the
period of one Cooper--pair ($n_x=1$).
 This periodicity is similar to the periodicity of the spectrum of
electrons on a lattice with respect to the addition of one flux quantum per
plaquette.  Thus, although the ratio
between the density of vortices and the density of charges in the
system is very small, the physically meaningfull ratio is the ratio of
 ${\overline n}_v$ to
$(n_x-[n_x])$ (where $[n_x]$ is the largest integer smaller than
$n_x$), and this ratio is not necessarily small.
Following this observation, {\it we limit ourselves from now on to the case
$0\le n_x<1$}.

Having described in detail the Josephson junction array to be considered, we
conclude this section by formulating precisely the question to be studied,
namely, {\it how do the physical properties of the array depend on the ratio
between the vortex density ${\overline n}_v$ and the charge density $n_x$?}
We start our examination of  that question by deriving an effective
action for the vortices in the array.

\vskip 1cm \noindent
{\large\bf 3. The effective action for the vortices}

The Hamiltonian (\ref{ham}) describes the Josephson junctions array
in terms of the sets of variables $\{n_i\},\{\phi_i\}$. In this section
we derive an equivalent description of the array in terms of the
vortex density $\rho^{vor}$, the vortex current ${\vec J}^{vor}$ and gauge
fields the vortices interact with. Our goals in attempting to derive
this description are three--fold. The first goal is to verify the
validity of our assertion that vortices are subject to Magnus force,
and that $n_x$ plays a role analogous to that of a magnetic field in
electronic dynamics.  The second goal
is to study the mutual interactions between vortices. The third goal is
to estimate the mass of the vortices. The  first two goals are
relatively easy to achieve. Estimating the mass of the vortex, however,
turns out to be a harder task, which we are able to handle only
approximately.

The effective action for vortices in a Josephson junction array was
first discussed by Eckern and Schmid, who considered the Hamiltonian
(\ref{ham}), with $n_x=0$. More generally, the effective action for
singularities in the phase configuration of a complex field was discussed
in various other contexts in physics. A particularly convenient method
for the derivation of such an action is the  "duality
transformation", developed and used by Jose {\it et.al.} \cite{Jose},
 Berezhinskii \cite{Berezhinskii},
Peskin \cite{Peskin},  Fisher and Lee \cite{Fisher}
 and others.
This method was applied to analyze the motion of
vortices in Josephson junction arrays (again, for the case
$n_x=0$) by Fazio, Geigenmuller and Schon \cite{Fazio}.

In our derivation of the effective action, we follow Fazio,
Geigenmuller and Schon\cite{Fazio}
by applying the
duality transformation to obtain an effective action for vortices
on a lattice. The action resulting from the duality transformation
(Eq. (\ref{action}) below) describes the vortices as bosons on a lattice
 interacting with an externally put vector potential, as well
as with a dynamical vector potential. The externally put vector
potential, which we denote by $\vec{\cal K}^{ext}$, satisfies
${\vec \nabla}\times
{\vec{\cal K}}^{ext}=2\pi\hbar n_x$.
The interaction with the dynamical vector potential mediates a
vortex--vortex interaction, composed of two parts. The first part is
the familiar logarithmic interaction. Its strength is proportional
to the Josephson energy $E_J$.
The second part, induced mostly by the charging energy of
the array,  is a short ranged velocity--velocity interaction.
The latter makes the vortices massive, since it includes a self interaction
term, quadratic in the vortex velocity. However, the mass defined by
this interaction is a "bare mass", that
 does not take into account
the periodic potential exerted on the vortices
by the  lattice. Generally speaking,
the periodic potential changes the bare mass into an effective band mass.
In an attempt to estimate the  band mass we write the
continuum limit of the vortices action. In the continuum language,
vortices are
massive particles interacting with an external vector potential,
with a periodic lattice potential and with one another.
Our analysis of this rather complicated dynamics follows the way the
dynamics of electrons on a lattice is analyzed. We start by neglecting
vortex--vortex interactions. We are then
faced with a single particle problem, in which a massive
vortex interacts with a
static vector potential ${\vec{\cal K}}^{ext}$, and with
a periodic lattice potential.
This problem  is identical to the
problem of an  electron under the effect of  a uniform magnetic field and
a lattice periodic potential, whose solution is well
known.
  When $n_x\ll 1$ the effect of the periodic potential can be
accounted for by changing the "bare mass" to an effective mass.
We limit ourselves to this case, and estimate the resulting
effective mass.  Then, we incorporate
the vortex--vortex interactions back into the action.

Before turning into the details of the derivation sketched in the last
paragraph, we pause to
define a notation.
We denote 3--vectors by bold--faced letters, and
their two spatial components by vector arrows. The electromagnetic
potential is then ${\bf A}=(A_0,A_x,A_y)=(A_0,{\vec A})$.
We number array sites by a subscript $i$.
The bond connecting a site $i$ to its neighbor on the right side
is denoted
by the subscript $i,x$. Similarly, the bond connecting the $i$'th site
to the site above it is denoted by the subscript $i,y$. The difference
operator $\vec\Delta$, a discretized version of $\vec\nabla$, is
defined accordingly. When operating on a scalar $\phi$, for example,
the $x$--component of $\vec\Delta_i$ is $\phi_j-\phi_i$ where $j$
is the neighbor to the right side of $i$.

Our derivation of the effective action starts by considering the
partition
function
\begin{equation}
{\cal Z}={\rm tr}e^{-\beta H}=\int D\{n_i\}\int
 D\{\phi_i\}e^{-{1\over\hbar}S(\{n_i(t)\},\{\phi_i(t)\})}
\label{parfunc}
\end{equation}
 where the action $S(\{n_i(t)\},\{\phi_i(t)\})\equiv
\int_0^\beta dt \left[ i\sum_i
\hbar n_i(t){\dot\phi}_i(t)-H(\{n_i(t)\},\{\phi_i(t)\})\right ]$
and the Hamiltonian is given by (\ref{ham}).
The path integral is to include all paths satisfying
 $n_i(\beta)=n_i(0)$ and $\phi_i(0)=\phi_i(\beta)$.
The variables $n_i$ are integers and therefore the path integral has to
be performed stepwize \cite{Swanson}. We limit ourselves to zero
temperature,
i.e., $\beta=\infty$.

The first step of the derivation follows closely previous
works \cite{Fazio}, and is
therefore given in  Appendix A.
Using the Villain approximation and the duality transformation method,
the path integral over the charge and phase degrees of freedom, $n_i$ and
$\phi_i$, is transformed to a path integral over a 3--component integer
 field  $\bf J^{vor}_i$ describing the vortex charge and density, and a
3--component real gauge
field $\bf {\cal K}_i$, to which $\bf J^{vor}_i$ is coupled. This gauge
field describes the charge degrees of freedom, to which it is related
through its derivatives.  The field
strengths associated with this gauge field,
 ${1\over{2\pi\hbar}}\epsilon^{\mu\nu\sigma}
\partial_\mu{\cal K}_{i,\nu}$ are  the Cooper--pairs
current and density on the $i$'th site.
 In terms of $\bf J^{vor}$ and
$\bf{\cal K}$, and in a gauge in which
${\vec\Delta}\cdot{\vec{\cal K}}=0$
the action is given by,
\begin{equation}
\begin{array}{ll}
S^{vor}=
\sum_i\left\{ i(\rho^{vor}_i-{{\overline n}_v}){\cal K}_{i,0}  +\
\ i{\vec J}^{vor}_i
\cdot({\vec{\cal K}}_i+{\vec{\cal K}}^{ext}_i)+{1\over{8\pi^2
 E_J}}[(\Delta_t{\vec{\cal K}}_i)^2
+({\vec\Delta}{\cal K}_{0i})^2]\right\}
 \\ \\
 +{{e^2}\over {2\pi^2\hbar^2}}{\sum_{ ij}}
({\vec\Delta}\times{\vec{\cal K}}_i)
{\hat C}^{-1}_{ij}
({\vec\Delta}\times{\vec{\cal K}}_j)
\end{array}
\label{action}
\end {equation}
where ${\vec\Delta}\times{\vec{\cal K}}^{ext}=2\pi\hbar n_x$.
This action describes the
vortices as bosons on a lattice,
interacting with an
externally put gauge field $\vec{\cal K}^{ext}$ whose spatial curl
is a constant, given by
$2\pi\hbar n_x$, and with a dynamical gauge field $\bf{\cal K}$. As
expected from
the similarity between Magnus and Lorenz forces, a moving vortex is
affected by the Cooper--pairs on the dots in the same manner a charged
particle is affected by a magnetic field. Moreover, the Josephson
currents between the dots affect the vortices in the same way
an electric field affects charged particles. The last three terms of the
action include the self energy of the field $\bf{\cal K}$. They are simply
understood once the relation between $\bf{\cal K}$ and the Cooper--pairs
currents and densities is taken into account.
The first two are the kinetic energy of the Josephson
currents (the transverse part of that current is
 ${{e}\over{\pi\hbar}}{\vec\Delta}{\cal K}_0$, and
${{e}\over{\pi\hbar}}\dot{\vec{\cal K}}$ is the longitudinal part).
The last term is
the charging energy (the  net charge
on the
$i$'th dot is ${1\over{2\pi\hbar}}{\vec\Delta}\times{\vec{\cal K}}_i$).
The transverse part of
the current satisfies a two dimensional Gauss law
${\vec\Delta}^2{\cal K}_0=4\pi^2E_J\rho^{vor}$ and mediates a
logarithmic interaction between the vortices. The excitation spectrum of
$\vec{\cal K}$ is  the spectrum of longitudinal oscillations of the
Cooper--pairs,
i.e., the plasma spectrum of the array.

Our next step is a formulation of the continuum limit of the action
(\ref{action}).
When doing that,
two points
should be handled carefully. The first is the periodic potential
exerted by the lattice on the vortex. This potential was studied in
detail by Lobb, Abraham and Tinkham\cite{Lobb}. Since Currents do not
flow uniformly within the array, the energy cost associated with a
creation of a vortex depends on the position of its center within a plaquette
(i.e., on the precise distribution of the currents circulating its core).
This energy cost is periodic with respect to a lattice spacing
 of the array, and
is independent of the sign of the vorticity.
The origin of this
potential can be visualized using the analogy with 2D electrostatics.
In that analogy,
a vortex is analogous to a charge in a two--dimensional world. A vortex on a
lattice is then analogous to a charge in a two dimensional world {\it in
which the dielectric constant varies periodically with position}. The
electrostatic energy of such a charge varies periodically with position,
too, and is independent of the sign of the charge. This energy cost
can then be interpreted as a periodic potential exerted by the lattice on the
vortices.
The characteristic  energy scale for that potential
is  $E_J$. Its
amplitude and functional form were  studied in Ref. \cite{Lobb}.
The amplitude was found to be $0.2E_J$ and $0.05 E_J$ for
square and triangular lattices, respectively.

A convenient way to incorporate the periodic dependence of the vortex potential
energy on the position of the vortex core within a plaquette is by
replacing the Josephson energy $E_J$ in the action (\ref{action}) by
a periodically space dependent function $\epsilon_J({\vec r})$,
that is non--zero
only along lattice bonds. The period of $\epsilon_J({\vec r})$ is obviously
the lattice spacing. The energy cost involved with the
Josephson currents then becomes $\int d{\vec r}
{1\over{8\pi^2
 \epsilon_J({\vec r})}}[({{\vec \nabla}}{\cal K}_{0}({\vec r},t))^2+
({\dot{\vec{\cal K}}}({\vec r},t))^2]$,
 and that energy cost
confines the currents to the lattice bonds. The effect of the spatial
dependence of $\epsilon_J({\vec r})$ on the interactions mediated by
${\cal K}_0,
{\vec{\cal K}}$ is discussed below.

The second point to be handled carefully when transforming to a
a continuum action is the short distance cut--off on the capacitance
matrix. The model we employ
does not attempt to describe statics and dynamics of
Cooper--pairs within superconducting dots. Thus, its continuum version
should not allow for excitations of
$\vec{\cal K}$ at wavelengths shorter than the lattice spaing.
This constraint is taken into account by introducing a high wave--vector
cut--off to the capacitance matrix, as it was done in Ref. \cite{Eckern}.

Taking into account the two points discussed above, the continuum limit of
the action (\ref{action}) is,
\begin{equation}
\begin{array}{ll}
S&=\int dt\int d{\vec r}\Bigg\{i[\rho^{vor}({\vec r},t)-{\overline n}_v]
{\cal K}_0({\vec r},t)+i{\vec J}({\vec r},t)\cdot [{\vec{\cal K}}
({\vec r},t)
+{\vec{\cal K}}^{ext}({\vec r})]\nonumber \\ \nonumber\\
&\hspace{2.5in}+{1\over{8\pi^2\epsilon_J({\vec r})}}[({\vec \nabla}
{\cal K}_0({\vec r},t))^2+
({\dot{\vec{\cal K}}}({\vec r},t))^2]
\Bigg\} \nonumber\\ \nonumber\\
&+{{e^2}\over{2\pi^2\hbar^2}}
\int dt \int d{\vec r}\int d{\vec r}'
\left[{\vec \nabla}\times{\vec{\cal K}}({\vec r},t)\right]
{\hat C}^{-1}({\vec r}-{\vec r}')
\left[{\vec \nabla}\times{\vec{\cal K}}({\vec r}',t)\right]
\label{conact}
\end{array}
\end{equation}

When the fields ${\cal K}_0, {\vec{\cal K}}$ are integrated out they mediate
mutual interactions between vortices and self interactions of a vortex
with itself. The spatial dependence of $\epsilon_J({\vec r})$ does not
significantly affect mutual interactions between vortices whose
distance is much larger than one lattice spacing. It does,
however, potentially affect the self interaction.

Consider  the action associated with a single vortex. As discussed above,
due to the spatial dependence of $\epsilon_J({\vec r})$,
 the interaction of the vortex with ${\cal K}_0$ yields a  periodically
space dependent potential energy \cite{Lobb}.
The coupling to $\vec{\cal K}$ results
in a kinetic energy.
To see that, note that the vortex current corresponding to a
moving vortex whose center is at ${\vec r}_0(t)$
is ${\vec J}^{vor}={\dot{\vec r}_0(t)}\delta\left({\vec r}-
{\vec r}_0(t)\right)$.
Substituting this expression in the action (\ref{conact}), we find
the part of the  action that depends on
 the vortex velocity, $\dot{\vec r}_0$, and the gauge field
it interacts with, $\vec{\cal K}$, to be
\begin{equation}
\begin{array}{ll}
& i\int dt {\dot{\vec r}_0(t)}\cdot{\vec{\cal K}}({\vec r}_0,t) \\ \\ +
&\int dt\Bigg\{\int d{\vec r}
{1\over{8\pi^2\epsilon_J({\vec r})}}({\dot{\vec{\cal K}}}({\vec r},t))^2
+{{e^2}\over{2\pi^2\hbar^2}}
 \int d{\vec r}\int d{\vec r}'\left[{\vec \nabla}\times{\vec{\cal K}}
({\vec r},t)\right]
{\hat C}^{-1}({\vec r}-{\vec r}')
\left[{\vec \nabla}\times{\vec{\cal K}}({\vec r}',t)\right]\Bigg\}
\end{array}
\label{svdyn}
\end{equation}
The gauge field $\vec{\cal K}$ can  be integrated out,
with the resulting
effective action for the vortex velocity $\dot{\vec r}_0$  being non--local
in time. However,
as pointed out by Eckern and Schmid \cite{Eckern} and by Fazio {\it et.al.}
\cite{Fazio}, the time non--locality can be neglected
as long as the characteristic frequencies involved in
${\dot{\vec r}_0(t)}$ are smaller than the Josephson plasma frequency
${1\over\hbar}\sqrt{8E_JE_C}$. This neglect is possible due to the gap in the
excitation spectrum of $\vec{\cal K}$, a gap that makes the time non--locality
short ranged \cite{Notea}. Having neglected the time non--locality, we find
the effective action for the vortex velocity $\dot{\vec r}_0$  to be,
\begin{equation}
 \int dt{1\over 2}m_{bare}{\dot{\vec r}_0(t)}^2
\label{baremass}
\end {equation}
where $m_{bare}$, the vortex bare mass, is defined by
 $m_{bare}\equiv{{\pi^2\hbar^2}\over{4E_C}}$. \cite{Eckern}
Thus, the interaction of the vortex with
$\vec{\cal K}$ results in a kinetic term.

Having integrated out both ${\cal K}_0$ and $\vec{\cal K}$ we have turned
the single
vortex action into an action of a charged particle interacting with a
periodic potential  and a magnetic field $2\pi\hbar n_x$.
For $n_x\ll 1$ the effect of the periodic
potential is to change the bare mass into an effective band mass.
Since the lowest band is the relevant one for bosons, the band mass
is always larger than the bare one \cite{Kittel}.
 The effective band mass
for $n_x=0$ was studied both theoretically and numerically by
Geigenmuller \cite{Geigenmuller} and by Fazio {\it
et.al.}\cite{Fazio} (see also references therein).
 While a qualitative estimate of the mass is easy to arrive at,
a quantitative determination depends on the precise details of the
periodic potential, and is therefore hard to obtain. Qualitatively,
the tight binding limit, in which  $E_J\gg E_C$,
is distinguished from the weak periodic potential limit, in which the
opposite condition applies. In the former, the effective band mass is
\begin{equation}
m_{band}\sim \hbar^2
\sqrt{{\alpha_1 }\over{E_JE_C}}e^{\sqrt{\alpha_2 {E_J\over E_C}
}}
\end{equation}
where $\alpha_{1,2}$ are numbers of order
unity \cite{Fazio}\cite{Geigenmuller}. In the latter,
\begin{equation}
m_{band}\sim  m_{bare}\left(1+({E_J\over E_C})^2\right )
\end{equation}
  The regime of parameters we
are interested in lies between the two limits. It is therefore
reasonable to assume that the band mass is larger than, but
of the order of, the bare one.

We now turn to discuss the many vortices configuration.
As we have argued above, the
discreteness of the array does not significantly affect
the mutual interaction
between vortices. Therefore, for the study of this interaction we
may replace $\epsilon_J({\vec r})$ by $E_J$.
 Then, the action
(\ref{conact}) can be written in momentum space as
\begin{equation}
\begin{array}{ll}
S=
\int dt\int {{d{\vec q}}\over{(2\pi)^2}}\left\{i[\rho^{vor}_{-{\vec q}}
-{\overline n}_v\delta({\vec q})]{\cal K}_{0{\vec q}}
+i{\vec J}^{vor}_{\vec q}\cdot({\vec{\cal K}}_{-{\vec q}}
+{\vec{\cal K}}^{ext}_{-{\vec q}})
+{1\over{8\pi^2E_J}}[|{\vec q}{\cal K}_{0{\vec q}}|^2+
|{\dot{\vec{\cal K}}}_{\vec q}|^2]
\right. \\ \\
+\left.{{e^2}\over {2\pi^2\hbar^2}}
|{\vec q}\times{\vec{\cal K}}_{\vec q}|^2{\hat C}^{-1}({\vec q})\right\}
\end{array}
\label{conactt}
\end{equation}
where $\rho^{vor}_{\vec q},
J^{vor}_{\vec q},{\vec{\cal K}}_{\vec q}, {\vec{\cal K}}^{ext}_{\vec q},
{\hat C}^{-1}({\vec q})$ are the
Fourier transforms of the corresponding quantities. The momentum space
representation used in Eq. (\ref{conactt}) is convenient for the
integration of the field $\bf{\cal K}$. The integration out of the time
component, ${\cal K}_0$, yields a density--density interaction between the
vortices, of the form $
{1\over 2}\int d{\vec q}{{E_J}\over{q^2}}|\rho^{vor}_{{\vec q}}
-{\overline n}_v\delta({\vec q})|^2$, where $q\equiv|{\vec q}|$.
When transformed back to real space,
this interaction is
\begin{equation}
{\pi E_J}\int d{\vec r}\int d{\vec r}'
(\rho^{vor}({\vec r})
-{\overline n}_v)\log{|{\vec r}-{\vec r}'|}(\rho^{vor}({\vec r}')-
{\overline n}_v)
\label{logint}
\end{equation}
Similarly, the integration of $\vec{\cal K}$ yields a current--current
interaction between the vortices.
In the gauge we use, $\vec{\cal K}_{\vec q}$ has only transverse
components. It
therefore mediates an interaction only between the transverse component of
the vortex currents.
Integrating out $\vec{\cal K}$, and neglecting again the slight
non--locality
in time, we find that the current--current interaction is given, in
momentum space, by
\begin{equation}
{\hbar^2\over{8 e^2}}\int {{d{\vec q}}\over{ q^2{\hat C}^{-1}(q)}}
|J^{vor}_{\perp{\vec q}}|^2
={\hbar^2\over{16 E_C}}
\int {{d{\vec q}}} |J^{vor}_{\perp{\vec q}}|^2
\label{curcur}
\end{equation}
where $J^{vor}_{\perp{\vec q}}\equiv
{{{\vec q}}\over q}\times{\vec J}^{vor}_{{\vec q}}$
is the transverse component of ${\vec J}^{vor}_{{\vec q}}$,
and the integral over
${\vec q}$ is cut--off at $q=2\pi$. The current--current interaction
is described in real space as,
\begin{equation}
{\hbar^2\over{16 E_C}}
\int d{\vec r}\int d{\vec r}'\int {{d{\vec q}}} [{\vec J}^{vor}
({\vec r})\times{\vec q}]
[{\vec J}^{vor}({\vec r}')\times{\vec q}] {1\over
{q^2}}e^{i{\vec q}\cdot ({\vec r}-{\vec r}')}
\label{velvel}
\end{equation}
As pointed out in Ref. \cite{Eckern},
this current--current interaction includes both a
self interaction term, that assigns a mass $m_{bare}$ to
each vortex, and a
velocity--velocity interaction between different vortices.
The former was discussed in the context of a single vortex. The latter
is short ranged. For large separations, it is
inversly proportional to the square of the distance between the
interacting vortices.

Eqs. (\ref{logint}) and (\ref{velvel}) both neglected the effect of
 the lattice structure of the array on the vortex--vortex
 interactions.  Similar
to the common practice in the analysis of electrons on a lattice, we
 assume that the sole effect of the lattice is to modify the
single vortex mass from the bare mass $m_{bare}$ to the effective band
mass $m_{band}$.
The current--current interaction in Eq. (\ref{velvel}) includes
a self interaction that assigns a mass
$m_{bare}$ to each vortex.
 To account for the modification of the mass by the lattice,
we add another
kinetic term to the action, of the form ${M^*}\int d{\vec r}
{{{\vec J}^{vor}({\vec r})^2}\over
{2\rho^{vor}({\vec r})}}$, where ${M^*}\equiv m_{band}-m_{bare}$.
Altogether, then, the
effective vortices action becomes,
\begin{equation}
\begin{array}{ll}
S^{vor}({\bf J^{vor}})=\int dt\Bigg\{&\int d{\vec r}\Big[
{M^*}{{{\vec J}^{vor}({\vec r})^2}\over
{2\rho^{vor}({\vec r})}}+i{\vec J}^{vor}({\vec r})\cdot
{\vec{\cal K}}^{ext}({\vec r})\Big]\\ \\
&+
{\hbar^2\over{16 E_C}}
\int d{\vec r}\int d{\vec r}'\Big[\int {{d{\vec q}}}
[{\vec J}^{vor}({\vec r})\times{\vec q}]
[{\vec J}^{vor}({\vec r}')\times{\vec q}] {1\over
{|{\vec q}|^2}}e^{i{\vec q}\cdot ({\vec r}-{\vec r}')}\\ \\
&+{{\pi E_J}}
(\rho^{vor}({\vec r})
-{\overline n}_v)\log{|{\vec r}-{\vec r}'|}(\rho^{vor}
({\vec r}')-{\overline n}_v)\Big]
\Bigg\}
\end{array}
\label{voracti}
\end{equation}

Eq. (\ref{voracti}) is a concise description of the dynamics
of the vortices,
since the only dinamical fields it includes are those of the
vortices.
It describes the vortices as interacting particles of mass
$m_{band}$ and
 an average
density ${\bar n}_v$, under the effect of a "magnetic field"
$2\pi\hbar n_x$. The vortices "filling factor" is then indeed
${\bar n}_v\over n_x$.

The current--current
interaction term in the action (\ref{voracti})  is somewhat
inconvenient for calculations.
Thus, in our analysis of the quantum  Hall fluid of vortices in the
next section we choose to reintroduce $\vec{\cal K}$, and
consider the vortices
as particles of mass ${M^*}$ interacting with a dynamical
vector potential $\vec{\cal K}$, as well as with
${\vec{\cal K}}^{ext}$ and
with one another.

We conclude this section with a few remarks regarding the dependence of
its results on the form of the capacitance matrix. The capacitance matrix
determines the bare mass of the vortex (see Eqs. (\ref{svdyn}) and
(\ref{baremass})) and the form of the vortex current--current interaction
(see Eq. (\ref{velvel})).
So far we considered a capacitance matrix that includes only nearest
neighbors coupling. The inverse capacitance matrix
describes then a two dimensional Coulomb interaction between
Cooper--pairs on the superconducting dots. In Fourier space, it is
proportional to $1\over q^2$.
Inclusion of capacitances to the ground and/or  capacitance
between dots that are not nearest neighbors result in a
screening of that interaction. Then, at small $q$,
${\hat C}^{-1}(q)\propto q^{-\alpha}$, with $0\le\alpha<2$.
This screening has two consequences.
First, the kinetic energy cost involved in a vortex motion,
i.e., its bare mass, is affected.
Second, the excitation spectrum of $\vec{\cal K}$ becomes gapless.
We now examine these consequences. Consider a vortex moving in a
constant velocity ${\vec v}$.
 As seen from Eq. (\ref{svdyn}), a
moving vortex acts like a source for the vector potential $\vec{\cal K}$.
The (imaginary time) wave equation for $\vec{\cal K}$ can be derived from
Eq. (\ref{svdyn}). In Fourier space its solution is,
\begin{equation}
{\cal K}_{\perp{\vec q},\omega}=iv_\perp
{{ \delta(\omega-{\vec q}\cdot{\vec v}})
\over{{\omega^2\over {4\pi^2 E_J}}+
{{e^2}\over{\pi^2\hbar^2}}q^2{\hat C}^{-1}(q)}}
\label{wave}
\end{equation}
where $v_\perp\equiv{{{\vec q}\times\vec v}\over q}$
is the transverse part of
the velocity vector. The longitudinal component of
$\vec{\cal K}$ vanishes in the
gauge we use.
The kinetic energy cost associated with the motion of a vortex is
the energy cost of the fields ${\dot{\vec{\cal K}}}$
and ${\vec \nabla}\times{\vec{\cal K}}$ it creates. It is
composed of two parts. The first, $\int {d{\vec r}}{1\over{8\pi^2E_J}}
{\dot{\vec{\cal K}}}^2$, is the kinetic energy cost
of the longitudinal
currents created by the motion of the vortex. The second,
${{e^2}\over{2\pi^2\hbar^2}}\int d{\vec r}\int d{\vec r}'
\left[{\vec \nabla}\times{\vec{\cal K}}({\vec r})\right ]
{\hat C}^{-1}({\vec r}-{\vec r}')\left
[{\vec \nabla}\times{\vec{\cal K}}({\vec r}')\right]$
is the cost in charging energy. A moving
vortex induces voltage drops between the superconducting dots, and those
result in a charging energy cost, determined by the capacitance matrix.
The first energy cost
is proportional to $v^4$, while the second is proportional to $v^2$.
Thus, the bare mass is determined by the charging energy.  Transforming
Eq. (\ref{wave}) to real space, and substituting into the expression for
the charging energy, we observe that the charging energy is finite as long as
$\alpha>0$, and diverges logarithmically with the system size for
$\alpha=0$.
The gapless excitations of $\vec{\cal K}$, characteristic of $\alpha<2$,
play a role when vortices accelarate or decelarate. Then, the coupling
of the vortices to these excitations (the "spin waves" \cite{Fazio})
becomes  a  weak mechanism for dissipation of a vortex kinetic
 energy \cite{Eckern}.
For the present context we note that the effect of a weak dissipative
mechanism on the quantized Hall effect was studied by Hanna and Lee
\cite{Hanna}. While some properties of the effect are affected by such a
mechanism, its main features are not.

\vskip 1cm \noindent
{\large\bf 4.  The quantum Hall fluid of vortices}
\vskip 0.5cm\noindent
{\large \bf 4.1 General discussion}

In the previous section we established the mapping of the vortex dynamics in
a  Josephson junction array on the problem of interacting charged
particles in a magnetic field.
We have also identified the ratio ${\overline n}_v\over n_x$
as the vortices filling factor. In this section we examine the formation
of QHE fluid state of vortices at appropriate filling factors.
While so far we have emphasized the
similarities between the dynamics of the vortices and that of electrons in
a magnetic field, in this section we must study the differences between the
two.
We begin by a general discussion of two of the differences. Then, we turn
 in the next subsection to a detailed calculation, using the
Chern Simon Landau Ginzburg
approach to the QHE, developed by Zhang, Hansson, Kivelson
and Lee \cite{Zhang}.

The first difference is in the statistics: while vortices are bosons,
electrons are fermions. This difference  changes the values
of the "magic" filling factors, and  eliminates the possibility of
a QHE in the absence of interactions. The filling factors at
which bosons form quantum Hall fluids are $p\over q$, where $p,q$
are integers, and one of them is even \cite{Read}. Fermi liquids of the
type discussed by Halperin, Lee and Read \cite{Halperinb} form at filling
factors $1\over(2n+1)$, where $n$ is an integer.

The second difference is in the interaction: the logarithmic interaction
between vortices is of longer range than the Coulomb interaction
between electrons. This difference leads to a modification
of the quantized Hall conductance, a modification of the
charge of Laughlin's quasiparticles, and, perhaps most interestingly,
to a modification of one of the diagonal elements of the linear
response function. While for a short range interaction these diagonal
elements  vanish in the long wavelength
low frequency limit $({\vec q},\omega\rightarrow 0)$, reflecting the
lack of longitudinal dc conductance in the QHE state, we find that the
logarithmic interaction makes one of the  diagonal elements non--zero.
In fact, rather than describing insulator--type
zero longitudinal response, as
expected from a QHE system, this element describes a
longitudinal response of the type usually associated with a
superconductor.
These consequences of the logarithmic interaction are all
derived in detail in the next subsection, where we also study the
difference between the linear response function and the conductivity.
 In this subsection
we preceed the derivation by a discussion of a thought experiment
that makes the role of the logarithmic interaction physically
transparent. The thought experiment we consider was extensively used
in the study of the Quantized Hall Effect, e.g., by Laughlin \cite{Laughlin}
 and Halperin \cite{Halperin},
and was proved very useful in understanding various aspects of the effect.

Consider a
"conventional" {\it electronic} quantum Hall system,
 in which a disk shaped
two dimensional electron gas (2DEG) is put in a strong magnetic field, and a
thin solenoid threads the disk at its center.
The flux through the solenoid is time dependent, and is denoted by
$\Phi(t)$.
If the electrons on the disk are in a QHE
state, the current density at any point is perpendicular to the {\it total}
electric field at that point. The time dependence of the flux induces an
electric field in the azymuthal direction, given by
${{e}\over{2\pi rc}}\dot\Phi(t)$, where $c$ is the speed of light, and $r$ is
the distance from the center.  Due to the
finite Hall conductance, this electric field induces a radial current, and,
consequently, a charge accumulation at the center of the disk. The
charge accumulated at the center during the interval $0<t<t_0$ is
given by ${{e\sigma_{xy}}\over{\Phi_0}}(\Phi(t_0)-\Phi(0))$
(where $\sigma_{xy}$ is the dimensionless
Hall conductivity and $\Phi_0\equiv {{hc}\over e}$ is the flux
quantum). This
charge accumulation, in turn, creates a radial electric field. Now, if
the electrons interact via a Coulomb interaction, the radial electric
field  is proportional to $1\over r^2$, i.e., it decays
faster than the azymuthal one. Then,   far away from the
center the electric field is predominantly azimuthal, and the currents
 are predominantly radial. However, if the electrons interact
 via a logarithmic interaction,
both the radial and azymuthal components of the
electric field are inversly proportional to $r$, and thus their
ratio is independent of $r$. The current then has both
radial and azimuthal components, and their ratio is
independent of $r$, too.
Moreover, the azymuthal component of the current is  proportional to the
flux at the center, and not to its time derivative.

The two components of the current and the charge accumulated
in the center can be  calculated using classical equations of
motion, since  in the absence
of impurities,
the classical and
quantum mechanical calculations coincide. Consider, therefore, the
hydrodynamical equation of motion of a
fluid of electrons
 in a magnetic field, whose
 electronic   density and velocity fields are denoted by
 $\rho({\vec r})$  and
${\vec v}({\vec r})$, respectively.  Assuming a uniform positive background
charge density $\overline\rho$ on the disk, this equation of motion is
\begin{equation}
m\rho({\vec r}){\dot{\vec v}}({\vec r})=
-\rho({\vec r}){\vec v}({\vec r})\times{\vec B}-
\int d{\vec r}'{\vec \nabla} V_{e-e}({\vec r}-{\vec r}')(\rho({\vec r}')-
{\overline\rho})\rho({\vec r})+
{{\dot\Phi}\over{2\pi r}}\rho({\vec r})
\label{conte}
\end{equation}
where ${\dot{\vec v}}({\vec r})$ is the complete time derivative of the
velocity field, $V_{e-e}$ is the electron--electron interaction
potential, and
we use a system of units where $e=c=1$. The
initial conditions corresponding to the scenario discussed in the previous
paragraph
 are $\rho({\vec r})={\overline\rho}$, ${\vec v}({\vec r})=0$ and
$\Phi(t=0)=0$.
 Due to the circular symmetry of both
Eq. (\ref{conte}) and its initial and boundary
conditions,  the current and density
 remain circularly symmetric when the flux is turned on,
 and the electron--electron interaction term
can be written as $V_{e-e}'(r)Q(r)\rho(r)$ where $Q(r)\equiv\int_0^r dr'
2\pi r'(\rho(r')-{\overline\rho})$ is the net charge within
a distance $r$ from the origin,
and $V_{e-e}'(r)\equiv{{\partial V_{e-e}(r)}
\over{\partial r}}$.
The conservation of charge constraint implies
${\dot Q}(r)=-2\pi r \rho(r)v_r(r)$, where $v_r$
is the radial component of $\vec v$.
The azymuthal  component of  Eq.  (\ref{conte}) can therefore
be written as,
\begin{equation}
2\pi r\rho(r){\dot v}_\phi={{\rho(r)}\over m}{\dot\Phi}-\omega_c {\dot Q}
\label{hydro}
\end{equation}
with $\omega_c\equiv{B\over m}$.
For values of $r$ far away from the center but not close to the edge the
density $\rho(r)$ remains approximately equal to $\bar\rho$ all along
the process, and thus $Q(r)$ is
$r$--independent. Within that approximation, and for such values of
$r$, the azymuthal equation can be
integrated and substituted in the radial one. The latter then becomes,
\begin{equation}
{\bar\rho}{\dot v}_r={{\omega_c^2Q}
\over {2\pi r}}-{{\bar\rho}
\over m}{{\omega_c\Phi}\over{2\pi r}}-{1\over m}V_{e-e}'(r)
Q(r){\bar\rho}+{1\over r}{\bar\rho}{\vec v}_\phi(r)^2
\label{radial}
\end{equation}
where the last term  is the centrifugal force.
Suppose now that the flux $\Phi$ is turned adiabatically on from zero to
$\Phi(t_0)$ in the interval $0<t<t_0$. For times $t\gg t_0$
the velocity field is purely azimuthal and  ${\dot v}_r=0$.
 Then,
if the potential
gradient ${\vec \nabla} V_{e-e}$ decays faster than $1\over r$,
so does also
the azymuthal velocity $v_\phi$, and
\begin{equation}
Q=\nu{{\Phi(t_0)}\over{\Phi_0}}
\label{coulomb}
\end{equation}
where $\nu={{\overline\rho}\over B}\Phi_0$ is the filling factor.
If $\Phi(t_0)=\Phi_0$ then the charge
accumulated near the origin is the charge of Laughlin's quasiparticle,
namely $-e\nu$.

However, in the case of a logarithmic interaction,
$V_{e-e}(r)=-V_0\log{r}$,
\begin{equation}
\begin{array}{ll}
v_\phi={\Phi\over{2\pi r m}}
{{V_0\nu}\over{\hbar\omega_c+V_0\nu}}\\     \\
Q ={\Phi\over\Phi_0}{\nu\over {1+{{ V_0\nu}\over{\hbar\omega_c}}  }  }
\end{array}
\label{loglog}
\end{equation}
The azymuthal current is then indeed inversly proportional to the
distance from
the origin, and  proportional to the flux $\Phi$.
In fact, the current is related to the vector potential created by
the solenoid, ${\vec A}^{sol}$, via a  London--type equation
\begin{equation}
{\vec \nabla}\times{\vec J}={{\bar\rho}\over m}
{{V_0\nu}\over{\hbar\omega_c+V_0\nu}}
{\vec \nabla}\times{\vec A}^{sol}
\label{london}
\end{equation}
 Thus, the  longitudinal response of the electrons
on the disk to the vector potential created by the solenoid
resembles  the longitudinal
response of a two dimensional superconducting disk in
a similar situation.

Two conclusions can be drawn from the above thought experiment. First, for
logarithmically interacting particles, the charge of the Laughlin
quasiparticle does not equal the filling factor, but  depends on the
interaction. And second, the transverse part of the linear response
function,
relating a transverse current to an externally applied
transverse vector potential, resembles that of
a superconductor. Since this response function is proportional to the
transverse current--current correlation function, the latter should
be expected to resemble a supeconductor, too. Both conclusions are
substantiated in the next subsection, and are applied to the study
of the quantum Hall fluid of vortices.

\vskip 1cm \noindent
{\large \bf 4.2 A study of the vortices QHE state by the Chern  Simon
Landau  Ginzburg approach}

In this subsection we use the Chern Simons Landau Ginzburg approach
to further analyze the properties of the quantized Hall fluid of vortices
formed at appropriate values of ${{{\overline n}_v}\over n_x}$. We start
by writing a Landau--Ginzburg action that describes the dynamics of
the vortices. We then perform a Chern--Simons singular gauge transformation
that attaches an even number of fictitious Cooper pairs to each vortex.
The resulting action, in which the order parameter describes
transformed "composite" bosons,
is convenient for a  saddle point analysis.
We find the uniform density saddle point that
describes a superfluid of composite bosons. By expanding the action
to quadratic order around that saddle point we  calculate the
response function of the vortices to an external probing field.  This
response
function, denoted by ${\hat\Sigma}$,
 is the ratio of the vortex density and current $\bf J^{vor}$
to an infinitesimal probing field $\bf{\cal K}^p$ applied externally
to the system.
The matrix ${\hat\Sigma}$ is calculated by
 adding an external probing field $\bf{\cal K}^p$
to the Lagrangian, and integrating out all
 the other fields to obtain an effective Lagrangian $L^{eff}$
in terms of $\bf{\cal K}^p$
only. The three components of $\bf J^{vor}$ are then
given by $J^{vor}_\alpha=-{{\partial
L^{eff}}\over {\partial {\cal K}^p_\alpha}}$,
and the  elements of ${\hat\Sigma}$ are \cite{Zhang}
\begin{equation}
\Sigma_{\alpha\beta}={{\partial{\bf
 J^{vor}_\alpha}}\over{\partial{\bf{\cal K}^p_\beta}}}
 \Bigg|_{{\bf{\cal K}^p_\beta}=0}=
- {{\partial\ }\over{\partial
{\bf{\cal K}^p_\alpha}}}{{\partial\ }\over
{\partial{\bf{\cal K}^p_\beta}}}L^{eff}
({\bf{\cal K}^p})\Bigg|_{{\bf{\cal K}^p}=0}
\label{sigmat}
\end{equation}
The physical meaning of ${\hat\Sigma}$, as well as
 the important distinction between the response to
$\bf{\cal K}^p$ and the response to the {\it total}
field $\bf{\cal K}^p+{\cal K}$, are
discussed after the  calculation is presented.

The Landau--Ginzburg action that describes the properties of
the vortices as they were found in section (3) is,
\begin{equation}
\begin{array}{ll}
S_{LG}({\tilde\psi},{\bf{\cal K}})
=
\int dt \Bigg\{&\int d{\vec r} \Big[\hbar{\tilde\psi^*}
\partial_t{\tilde\psi}+
{1\over {2{M^*}}}
|(i\hbar{\vec \nabla}-{\cal K}^{ext}-{\cal K}){\tilde\psi}|^2+
{1\over{8\pi^2E_J}}{\dot{\vec{\cal K}}}^2\Big]\\ \\
&+\int d{\vec r}\int d{\vec r}'
\Big[\pi E_J(|{\tilde\psi}({\vec r})|^2-{\overline n}_v)
\log|{\vec r}-{\vec r}'|(|{\tilde\psi}({\vec r}')|^2-{\overline n}_v)\\ \\
&+{{e^2}\over {2\pi^2\hbar^2}}
[{\vec \nabla}\times{\vec{\cal K}}({\vec r})]{\hat C}^{-1}({\vec r}-{\vec r}')
[{\vec \nabla}\times{\vec{\cal K}}({\vec r}')]\Big]\Bigg\}
\end{array}
\label{lg}
\end{equation}
The fields ${\tilde\psi},{\bf {\cal K}^{ext}},
{\bf{\cal K}}$ all depend on ${\vec r}$ and on
$t$. For the brevity of the expressions we omit this dependence whenever
this omission does not lead to confusion. The field $\tilde\psi$, the
order parameter for the vortices, satisfies bosonic commutation relations.

Restricting our derivation to the "fundamental" fractions $1\over\eta$,
where $\eta$ is an even number, our first step in analysing the action
(\ref{lg}) is the
Chern  Simon  singular gauge transformation, in which the field
${\tilde\psi}({\vec r},t)$ is transformed to
\begin{equation}
\psi({\vec r},t)=e^{i\eta\int d{\vec r}'{\rm arg}({\vec r}-{\vec r}')
|{\tilde\psi}({\vec r}',t)|^2}\ \ {\tilde\psi}({\vec r},t)
\end{equation}
where ${\rm arg}({\vec r}-{\vec r}')$ is the angle the vector
${\vec r}-{\vec r}'$ forms with
the $x$ axis.
Since $\eta$ is an
even integer  the field $\psi$ has the same statistics as $\tilde\psi$, i.e.,
bosonic.
Note that $|{\tilde\psi}({\vec r},t)|=|\psi({\vec r},t)|$. The singular gauge
 transformation shifts the phase of the field. Denoting the phase of
$\psi({\tilde\psi})$ by $\theta
({\tilde\theta})$, the Chern--Simons transformation amounts to
${\vec \nabla}\theta({\vec r},t)={\vec \nabla}
{\tilde\theta}({\vec r},t)-{1\over\hbar}{\vec a}({\vec r},t)$ where
${\vec a}({\vec r},t)\equiv \hbar\eta\int d{\vec r}'
{{{\hat z}\times({\vec r}-{\vec r}')}\over
{|{\vec r}-{\vec r}'|^2}}|{\tilde\psi}({\vec r}',t)|^2$.
The Chern--Simons field ${\vec a}$ has a gauge freedom, which we fix below.
 Writing  $\psi({\vec r},t)\equiv
\sqrt{n_v({\vec r},t)}e^{i\theta({\vec r},t)}$, the
 above Landau--Ginzburg functional  becomes
\begin{equation}
\begin{array}{lll}
S_{LG}(n_v,\theta,{\bf{\cal K}},{\bf a})
=
\int dt \Bigg\{\int d{\vec r} \Big[in_v(\hbar\partial_t\theta -a_0)+
{n_v\over {2{M^*}}}(\hbar{\vec \nabla}\theta-{\cal K}^{ext}-
{\cal K}+{\vec a})^2+ \\ \\
 {\hbar^2\over {2{M^*}}}({\vec \nabla}\sqrt{n_v})^2+
{1\over {8\pi^2E_J}}{\dot{\vec{\cal K}}}^2+{i\over{4\pi\eta\hbar}}
\epsilon^{\mu\nu\sigma}a_\mu\partial_\nu a_\sigma\Big]\\
 \\
+\int d{\vec r}\int d{\vec r}'\Big[\pi E_J(n_v({\vec r})-
{\overline n}_v)\log|{\vec r}-{\vec r}'|
(n_v({\vec r}')-{\overline n}_v)\\ \\
+{{e^2}\over {2\pi^2\hbar^2}}[{\vec \nabla}\times
{\vec{\cal K}}({\vec r})]{\hat C}^{-1}({\vec r}-{\vec r}')
[{\vec \nabla}\times{\vec{\cal K}}({\vec r}')]\Big]\Bigg\}
\end{array}
\label{lgcs}
\end{equation}
While the Euler--Lagrange equations of motion for the original field
$\tilde\psi$ required its phase to be multiply valued  (for
a minimization of the kinetic energy), the equations of motion for the
transformed field $\psi$ allow for a solution in which the magntiude and
the phase of the field
are constant. It is straight forward to see that the action (\ref{lgcs})
is minimized when,
\begin{equation}
n_v({\vec r},t)={\overline n}_v
\label{nv}
\end{equation}
\begin{equation}
{\vec \nabla}\theta({\vec r},t)=0
\label{theta}
\end{equation}
\begin{equation}
{\vec a}({\vec r},t)={\vec{\cal K}}^{ext}({\vec r})
\label{va}
\end{equation}
\begin{equation}
{\cal K}({\vec r},t)=0
\label{sa}
\end{equation}
This minimum of the action describes a state in which the vortex
density is
constant
on the average  and
the Chern--Simons field ${\vec a}$
cancels  ${\cal K}^{ext}$ on the average. We now expand
the action around these minimum values. Around the saddle point the
phase $\theta$ is singly valued, and thus we can choose a gauge in which
 $\theta=0$ identically, and the field $\psi$ is real.
Writing, then, $n_v={\overline n}_v+\delta n_v$,
$\psi=\sqrt{{\overline n}_v}+{{\delta n_v}\over{2\sqrt{{\overline n}_v}}}$
and ${\vec a}={\vec{\cal K}}^{ext}+{\delta{\vec a}}$, we find that
the quadratic deviations from
the extermum point (\ref{nv})--(\ref{sa}) are described by the
following action,
\begin{equation}
\begin{array}{ll}
S_{LG}(\delta n_v,{\bf{\cal K}},{\bf a})\approx\int dt\int d{\vec r}
[ia_0\delta{n_v}+{{\bar n}_v\over {2{M^*}}}({\vec{\cal K}}+{\delta{\vec a}})^2+
{\hbar^2\over {8{M^*}}}{{({\vec \nabla}\delta{n_v})^2}\over{{\overline n}_v}}+
{1\over{8\pi^2 E_J}}{\dot{\vec{\cal K}}}^2
+{i\over{4\pi\eta\hbar}}
\epsilon^{\mu\nu\sigma}a_\mu\partial_\nu a_\sigma] \\
 \\
+\int dt\int d{\vec r}\int d{\vec r}'\Big[\pi E_J\delta n_v({\vec r})
\log|{\vec r}-{\vec r}'|
\delta n_v({\vec r}')
+{{e^2}\over {2\pi^2\hbar^2}}({\vec \nabla}\times{\vec{\cal K}}({\vec r}))
{\hat C}^{-1}({\vec r}-{\vec r}')
({\vec \nabla}\times{\vec{\cal K}}({\vec r}'))\Big]
\end{array}
\label{quad}
\end{equation}

For the calculation of the conductivity matrix and the correlations
functions, we  imagine coupling the vortices to an
infinitesimal  3--vector probing field
$\bf{\cal K}^p$. Naturally, linear response functions are more conveniently
described in Fourier space. Thus, we write the action (\ref{quad})
in the presence of the probing field $\bf{\cal K}^p$, in Fourier space, as
\begin{equation}
\begin{array}{ll}
S_{LG}(\psi,{\bf{\cal K}},{\bf a},{\bf {\cal K}^p})\approx\int {{d\omega}
\over{2\pi}}\int
 {{d{\vec q}}\over{(2\pi)^2}}
&\Bigg[i(a_0({\bf q})+{\cal K}^p_0({\bf q}))
\delta{n_v}(-{\bf q})+{{\bar n}_v\over {2{M^*}}}|{\vec{\cal K}^p}({\bf q})+
{\vec{\cal K}}({\bf q})+{\delta{\vec a}}({\bf q})|^2\\ \\ & +
{\hbar^2\over {8{M^*}}}{{|{\vec q}\delta{n_v}({\bf q})|^2}
\over{{\overline n}_v}} +
{1\over{8\pi^2E_J}}|\omega{\vec{\cal K}}({\bf q})|^2
+{i\over{4\pi\hbar\eta}}
\epsilon^{\mu\nu\sigma}a_\mu({\bf q})q_\nu a_\sigma({\bf q})\\
 \\ &
+{{2\pi^2 E_J}\over{{\vec q}^2}}|\delta n_v({\bf q})|^2
+{{e^2}\over{2\pi^2\hbar^2}}|{\vec q}\times{\vec{\cal K}}({\bf q})|^2
{\hat C}^{-1}({\vec q})\Bigg]
\end{array}
\label{quadq}
\end{equation}
where ${\bf q}_0\equiv\omega$. We choose the Coulomb gauge for
$\bf{\cal K}^p$, i.e.,
${\vec q}\cdot{\vec{\cal K}^p}({\vec q})
=0$. Thus, the probing field becomes a 2--component vector.
Since the three components of $\bf J^{vor}$ are constrained by the conservation
of vorticity, $\bf J^{vor}$ is  effectively a two--component vetor, too,
and $\Sigma_{\alpha\beta}$ is a $2\times 2$ matrix. The indices
 $\alpha$ and
$\beta$ take the
values $0$ (for the time component) and $\perp$ (for the component
perpendicular to ${\vec q}$).
 The
integration of the fields $\delta n_v,{\vec {\cal K}},{\bf a}$
is easily carried out, since (\ref{quadq}) is quadratic in
all fields. The resulting effective Lagrangian is,
\begin{equation}
L^{eff}({\cal K}^p)=
\frac{
{{({\cal K}^p_\perp)^2}\over 2}\Big [{ {{{M^*}\omega^2}\over
{2\pi\hbar\eta\bar n_v}}+
{{2\pi E_J}\over{\hbar\eta}}+{{\hbar q^4}\over{8\pi{M^*}{\bar n_v}\eta}}
}\Big ]
+{{q^2({\cal K}^p_0)^2}\over 2 D}-
{i q{\cal K}^p_0{\cal K}^p_\perp}
}
{  {{{M^*}\omega^2}\over{{D}\bar
n_v}}+{4\pi^2 E_J\over {D}}+2\pi\hbar\eta
+{{\hbar^2 q^4}\over{4{M^*}{\bar n_v}D}}}
\label{eleff}
\end{equation}
where  ${D}\equiv 2\pi\hbar\eta
\left[{{M^*}\over{\bar n_v}}+{1\over{
{e^2\over{\pi^2\hbar^2}}
q^2{\hat C}^{-1}(q)+{{\omega^2}\over {4\pi^2E_J}} }}\right]^{-1}$.
In the limit of $q,\omega\rightarrow 0$ and for ${\hat C}(q)=C_{nn}q^2$
$D=2\pi\hbar\eta
\left[{{M^*}\over{\bar n_v}}+2m_{bare}\right]^{-1}$.

Eq. (\ref{sigmat}) expresses the matrix elements of ${\hat\Sigma}$ in
terms of second derivatives of this effective Lagrangian with
respect to ${\cal K}_0^p$ and ${\cal K}_\perp^p$.
Each of the four components of the matrix $\Sigma_{\alpha\beta}$
warrants a short discussion.  First,
the transverse component of the vortex current and the transverse
component of the gauge field $\bf{\cal K}^p$ are related, in the limit
${\vec q},\omega\rightarrow 0$, by
\begin{equation}
J^{vor}_\perp=-{ {2\pi E_JD}\over{ 4\pi^2\hbar\eta E_J+
2\pi\hbar^2\eta^2 D}}{\cal K}^p_\perp
\label{perp}
\end{equation}
This London--type of relation was anticipated by Eq. (\ref{london}).
It is characteristic of superconductors, and
is very different from the insulating behaviour characteristic of
the diagonal components of the response functions in QHE systems.
This difference results from the static vortex--vortex interaction being
of a long range.
Defering the discussion of the effect of Eq. (\ref{perp}) on the
longitudinal conductivity to a later stage, we now point out its
effect on the vortex current--current correlation function. By
the fluctuation--dissipation theorem,
\begin{equation}
\begin{array}{l}
\langle{ J^{vor}_\perp
 J^{vor}_\perp}\rangle_{{\vec q},\omega}=
 {\rm Im}\Sigma_{\perp\perp}({\vec q},\omega)
 \\     \\
 \int d\omega'{\rm P}({1\over{\omega-\omega'}})\langle{
 J^{vor}_\perp J^{vor}_\perp}\rangle_{{\vec q},\omega'}
 ={\rm Re}\Sigma_{\perp\perp}({\vec q},\omega)
\end{array}
\label{fdkk}
\end{equation}
where $\rm P$ denotes the principal part of the integral, and the second
line is an application of Kramers--Kronig relations \cite{Forster}.
For an insulator,
${\rm Re}\Sigma({\vec q},\omega)\propto\omega^2$ when
$\omega\rightarrow 0$.
This is also the case for QHE systems with short range
interactions \cite{Zhangb}.
For a superconductor ${\rm Re}{\hat\Sigma}({\vec q},\omega)$ approaches a
constant in the $\omega\rightarrow 0$ limit. As we now see,
so is also the case for
a QHE system in which the interactions are logarithmic.
In the particular problem we study, this constant is
$-{ {2\pi E_JD}\over{4\pi^2\hbar\eta  E_J+2\pi\hbar^2\eta^2 D}}$.

Second, we note that the compressibility of the vortex fluid vanishes
in the limit ${\vec q},\omega\rightarrow 0$,
as is manifested by the absence of low frequency poles in the
density--density correlation function. Like its electronic analog,
the quantum Hall fluid of vortices is incompressible.

Third,  the Hall component of the linear response function is
given, in the limit ${\vec q},\omega\rightarrow 0$, by,
\begin{equation}
\Sigma_{0,\perp}={{-iq}\over{{{4\pi^2 E_J}\over D}+2\pi\hbar\eta}}
\label{hall}
\end{equation}
If the vortex--vortex interaction was of shorter range, the long wavelength
limit of $\Sigma_{0,\perp}$ would satisfy
$\Sigma_{0,\perp}={{iq}\over{2\pi\hbar\eta}}$, seemingly demonstrating the
quantization of the Hall conductivity \cite{Zhang}. We further comment on
the difference between the two expressions below.

The qualitative effect the logarithmic interaction has on the transverse and
Hall components of the linear response function raises the following question:
 does the conductivity of the system we study have the
properties of the conductivity matrix of a QHE system, namely,
zero longitudinal conductivity  and quantized Hall conductivity?
To answer this question, we clarify
 the relation between the response function
$\Sigma_{\alpha\beta}$ and
the vortex conductivity matrix.
A similar  relation was discussed, in the context of
the QHE, by Halperin \cite{Halperin},
Laughlin \cite{Laughlinb}, Halperin Lee and Read \cite{Halperinb} and Simon
and Halperin \cite{Simon}.
The transport of vortices in the array is probed
by externally applied (number)
  density and current of Cooper--pairs, given, respectively, by
${1\over{2\pi\hbar}}{\vec \nabla}\times{\cal K}^p$,
and $-{1\over{2\pi\hbar}}{\vec \nabla}{\cal K}_0^p-{1\over{2\pi\hbar}}
{\dot{\vec{\cal K}^p}}$.
The matrix $\hat\Sigma$ is defined such that
${\bf J^{vor}}={\hat\Sigma}{\bf{\cal K}^p}$. However, the
vortices themselves contribute to the Cooper--pair  density and current,
with the most trivial contribution being the circulation of current
around each vortex center. The {\it total} Cooper--pair density and current
are therefore given by the derivatives of a total gauge field, composed of
the probing field $\bf{\cal K}^p$ and the field induced by the vortex
density and
current, denoted by
 $\bf{\cal K}^{ind}$. The latter is proportional to the vortex density and
current $\bf J^{vor}$. Thus, we can define a matrix $\hat V$ such that
\begin{equation}
{\bf{\cal K}^{ind}}\equiv{\hat V} {\bf J^{vor}}={\hat V}{\hat\Sigma}
{\bf {\cal K}^p}
\label{vmat}
\end{equation}
Consequently, the total field is ${\bf {\cal K}^{tot}}=
(1+{\hat V}{\hat\Sigma}) {\bf {\cal K}^p}$, and
\begin{equation}
{\bf J^{vor}}={\hat\Sigma}(1+{\hat V}{\hat\Sigma})^{-1}{\bf{\cal K}^{tot}}
\label{aaa}
\end{equation}
Thus, the  matrix ${\hat\Sigma}(1+{\hat V}{\hat\Sigma})^{-1}$
relates ${\bf {\cal K}^{tot}}$ to
the vector $\left (\begin{array}{c}\rho^{vor}
\\ J_\perp^{vor}
\end{array}\right )$.
The vortex conductivity matrix $\sigma^{vor}$,  relating the vortex
current $\left(\begin{array}{c}J_{||}^{vor}\\ J_\perp^{vor}
\end{array}\right )$
 to the {\it total driving force
vector}
$\left(\begin{array}{c}-i{{\vec q}\over{2\pi\hbar}}{\cal K}_0^{tot} \\ -
i{\omega\over{2\pi\hbar}}{\vec{\cal K}}^{tot}\end{array}\right )$ is then,
\begin{equation}
\sigma^{vor}=\left(\begin{array}{ll}
             -{\omega\over q}& 0 \\
                    0 & 0
                    \end{array}\right)
                    {\hat\Sigma}(1+{\hat V}{\hat\Sigma})^{-1}
             \left(\begin{array}{ll}
              i{{2\pi\hbar}\over q} & 0\\
                     0& i{{2\pi\hbar}\over\omega}
\end{array}\right)
\label{condu}
\end{equation}
where the leftmost matrix converts $\rho^{vor}$ to $J^{vor}_{||}$.

Eq. (\ref{condu}) defines the vortex conductivity matrix in terms of the
matrices ${\hat\Sigma}$ and $\hat V$. The matrix ${\hat\Sigma}$
is defined by Eqs.
(\ref{sigmat}) and (\ref{eleff}).
The matrix $\hat V$, relating $\bf J^{vor}$ to ${\cal K}^{ind}$, is specified
by the action (\ref{quadq}) to be
\begin{equation}
{\hat V}=\left ( \begin{array}{cc}
 {{4\pi^2 E_J}\over q^2} & 0 \\ \\
           0 & 1\over{ {\omega^2\over {4\pi^2E_J}}+{e^2\over{\pi^2\hbar^2}}
q^2{\hat C}^{-1}(q)}
\end{array}
 \right )
\label{vvv}
\end{equation}
The upper left element describes the field ${\cal K}_0$ created by a vortex
density $\rho^{vor}$. The gradient of that field is the transverse
current circulating around the vortex center. The bottom right element
describes the field ${\cal K}_\perp$ created by a transverse vortex current
$J^{vor}_\perp$, and is obtained from Eq. (\ref{quadq}) by taking its
derivative with respect to ${\cal K}_\perp({\bf q})$.

Substituting the matrices $\hat\Sigma$ and $\hat V$ to
Eq. (\ref{condu}), we find, to leading order in $q,\omega$,
\begin{equation}
{\hat\sigma}^{vor}=\left( \begin{array}{ll}
i{{\omega{M^*}}\over{\eta^2 n_v}} & 1\over\eta \\ \\
-{1\over\eta} & -i{{\omega{M^*}}\over{\eta^2 n_v}}
\end{array}\right )
\label{almost}
\end{equation}
In the limit ${\vec q},\omega\rightarrow 0$ the diagonal terms vanish, and the
vortex current satisfies
\begin{equation}
\vec J^{vor}=-{i\over{\eta 2\pi\hbar}}{\hat z}\times({\vec q}{\cal K}_0+
\omega{\vec{\cal K}})
\label{sofsof}
\end{equation}
Eqs. (\ref{almost}) and (\ref{sofsof}) describe a quantized Hall effect:
the current is purely perpendicular to the total "driving force", and
the Hall conductivity
is quantized. Contrasting Eqs. (\ref{perp}) and (\ref{hall})
with Eq. (\ref{sofsof}) we can finally summarize the effect of
the logarithmic interaction on the linear response of the system:
 the dc
conductivity, which is the $q,\omega\rightarrow 0$
 response to the {\it total} driving force,
has the usual form of the quantum Hall conductivity, and is unaffected by
the interaction. The correlation functions, on the other hand, determined
by the response to the {\it externally applied} driving force, are affected
by the interactions even in the $q,\omega\rightarrow 0$
 limit, with the most notable effect
being on the transverse current--current correlation function.

We conclude this section by relating the vortices conductivity,
 calculated above, to the electric conductivity and resistivity,
 which are
the quantities typically measured in experiments. The electric conductivity is
the matrix relating voltage drops (or, in the continuum limit, electric
fields) between superconducting dots
to the electric Josephson current flowing in the array. The electrostatic
potential at a point ${\vec r}$ is given by ${{e}\over{\pi\hbar}}
\int d{\vec r}' {\hat C}^{-1}({\vec r}-{\vec r}')
{\vec \nabla}\times{\vec{\cal K}}({\vec r}')$.
Thus, in Fourier space the $\vec q$ component of the
electrostatic potential is $i{{e}\over{\pi\hbar}}
q{\hat C}^{-1}(-{\vec q}){\cal K}_\perp({\vec q})$ and the
{\it longitudinal}
electric field is $-{{e}\over{\pi\hbar}}
q^2{\hat C}^{-1}(-{\vec q}){\cal K}_\perp({\vec q})$. Now, by deriving an
equation of motion
for ${\cal K}_\perp$ from
the action (\ref{conactt}), we see that a $dc$ {\it transverse} vortex current
${\vec J}^{vor}_\perp$
creates a field
${\vec{\cal K}}_\perp$ given by
\begin{equation}
{\vec J}^{vor}_{\perp{\vec q}}={{e^2}\over{\pi^2\hbar^2}}
q^2{\hat C}^{-1}({\vec q}){\cal K}_\perp({\vec q})
\end{equation}
i.e., a transverse
vortex current ${\vec J}^{vor}_{\perp{\vec q}}$ induces a longitudinal
electric field
 ${{\pi\hbar}\over{e}}{\vec J}^{vor}_{\perp{\vec q}}$.
A similar argument regarding the relation of the longitudinal
vortex current to the transverse electric field leads to the
conclusion that a vortex current ${\vec J}^{vor}_{{\vec q}}$ creates an
electric field ${{\pi\hbar}\over{e}}{\hat z}
\times{\vec J}^{vor}_{\perp{\vec q}}$.
The Josephson charge current, on the
other hand, is $-i{e\over{\pi\hbar}}{\hat z}\times
({\vec q}{\cal K}_0+{\omega{{\cal K}_\perp}})$, i.e., it is proportional
to the "driving force" acting on the vortices. Thus, the matrix relating
the Josephson current to the electric field is proportional to the matrix
relating the driving force acting on the vortices to the vortex current,
or, explicitly,
\begin{equation}
\rho^{el}={2\pi\hbar\over{(2e)^2}}\sigma^{vor}
\end{equation}
where $\rho^{el}$ is the electric {\it resistivity} matrix of the array
\cite{Choi}. This result can be simply concluded from Eq. (\ref{sofsof}).
The right hand side of that equation is the Josephson current, divided
by $2e$. The left hand side is proportional and perpendicular to the electric
field. The electric field  is then proportional and perpendicular to
the Josephson current, with the proportionality constant being ${{2\pi\hbar}
\over{(2e)^2\eta}}$.
The quantum Hall fluid of vortices manifests itself in electronic
properties of the array -- the longitudinal electric resistivity vanishes,
and the Hall electric resistivity is quantized.

\vskip 1cm

\noindent
{\large\bf 5. Conclusions}

In the previous sections we presented a study of the transport of
vortices in an array of Josephson junctions described by the
Hamiltonian (\ref{ham}). In particular, we focused on a quantum
Hall fluid formed by the vortices at appropriate values of
${\bar n}_v\over n_x$. In this section we summarize the results of
this study, and comment on a few open questions.

Our study was motivated by the analogy between Magnus force acting
on vortices and Lorenz force acting on charges in a magnetic field.
In this analogy, fluid density plays a role analogous to a magnetic
field, and fluid current density plays a role analogous to an
electric field. Quantum mechanics extends the analogy further:
a fluid particle is found to play a role analogous to that
of a flux quantum. This analogy motivates the search for a quantized
Hall effect for the vortices. The vortices' filling factor is
identified with the ratio of the vortex density to the fluid density.
This ratio is very small  for superconducting films, and it is
this smallness that motivates the study of the Josephson junction array.
Due to the periodicty of the spectrum of the Hamiltonian (\ref{ham})
with respect to the parameter $n_x$, the effective filling factor
becomes ${\bar n_v}\over {n_x(mod 1)}$, which can be made of order unity.

The dynamics of the vortices in a Josephson junction array was
studied in section (3). It is found to be that
of massive interacting charged particles under the effect of a magnetic field
and a periodic potential.  The
magnetic field is $2\pi\hbar n_x$.
The effect of the periodic potential is taken into account in
an effective mass approximation, changing the mass from a
bare mass to an effective band mass. The effective
mass is exponentially large for
$E_J\gg E_C$, and of the order of ${{\pi^2\hbar^2}\over{4E_C}}$ for
$E_J\widetilde{>}E_C$. Being interested in a phenomenon resulting
from a motion of vortices, we obviously consider the latter
regime. The mutual interaction between vortices consists of a
velocity independent logarithmic interaction, whose strength is
proportional to $E_J$,  and a short ranged
velocity--velocity interaction.

In view of the mapping  of the dynamics of the vortices on that of
massive interacting charged particles in a magnetic field, the existence
of a quantum Hall fluid phase is to be expected. In Section (4) we
examine some of the properties of that phase, but we leave
unexplored some other important propoerties. Most notable among the
latter are the regime of vortices filling factors at which
the quantum Hall fluid is the lowest energy state, and the energy
gap for excitations above that fluid.

Our study of the quantum Hall fluid is performed by means of the
Chern Simon Landau Ginzburg approach to the quantum Hall effect.
When ${{\bar n}_v\over n_x}={1\over \eta}$ with $\eta$ being an even
 integer, the vortices Landau--Ginzburg action is found to have
a saddle point corresponding to
a quantum Hall fluid. By hierarchical construction such saddle points
can be found for ${{\bar n}_v\over n_x}={p\over q}$, with $p,q$ being
one even and one odd integer.
The properties of the vortices quantum Hall fluid are studied
 within a quadratic expansion of the action around the corresponding
saddle point. We find that the vortices conductivity matrix  shows a
typical QHE behavior, i.e., zero diagonal elements and quantized
non--diagonal elements. However, we find the
${\vec q},\omega\rightarrow 0$ limit of the current--current correlation
functions
in the ground state to be different from those of a typical quantum
Hall  state,
due to the long range logarithmic interaction. In particular,
the transverse current--current correlation function is predicted
to bahave like that of a superconductor, rather than an insulator.
 For large arrays (larger than an
effective London penetration length) the vortex--vortex interaction is
screened.
Then, both the conductivity matrix and the correlation functions are
expected to behave, in the $dc$ limit, like those of a typical quantum
Hall state.

A necessary condition for the quantum Hall fluid to be the lowest energy
state is presumably that the ground state at $n_x=0$ and ${\bar n}_v\ne 0$
(infinite filling factor) is a superfluid of vortices, i.e., an insulator.
The observation, by van der Zant {\it et. al.}
\cite{Zant}, of a magnetic field tuned transition points at
the regime of parameters
in which this condition is satisfied, namely,
$E_J\approx E_C$ and $0.3> {\bar n}_v>0.15$. In this regime of
parameters we expect the quantum Hall fluid to be the ground state at
large filling factors, and the Abrikosov lattice to be the ground state
at small filling factors.
 This expectation is based on the phase diagram
of a two dimensional electron gas. For the latter, if the ground state
at zero magnetic field is a Fermi liquid,  then the ground state
at large filling factors $(\widetilde{>}0.2)$  is the quantum Hall fluid
and the ground state at low filling factors is the Wigner lattice.

\vskip 1cm
\noindent
{\large\bf Appendix A -- The Villain approximation and the
duality transformation}

The starting point of this appendix is the expression of the
partition function as a path integral over the phase and number
sets of variables $\{\phi_i\},\{n_i\}$, Eq. (\ref{parfunc}).
Using the Villain approximation and the duality transformation we
transform that path integral to a path integral over an integer
3--component vector field $\bf J^{vor}$, describing the vortex
density and current, and a real 3--component vector gauge field,
$\bf{\cal K}$. The action in terms of $\bf J^{vor}$ and $\bf {\cal K}$, to
be derived below, is given by Eq. (\ref{action}). The following derivation
follows the method of Fazio {\it et.al.}\cite{Fazio}

In the Villain approximation the imaginary time integral is done in
discrete steps, where the size of each step,  denoted by $\tau_0$, is
of the order of the inverse Josephson plasma frequency
$\omega_J\equiv \hbar^{-1}\sqrt{8E_JE_C}$.
 Each term in the Josephson energy part of
the path integral is approximated by a Villain form (we put $\hbar=1$
throughout the appendix, and restore its value in the final formula),
\begin{equation}
\begin{array}{ll}
e^{-\tau_0 E_J (1-\cos(\phi_{ij}-A_{ij}))}
\approx\sum_{v_{ij}=-\infty}^{\infty}&e^{-{1\over 2}\tau_0 E_J(\phi_{ij}-A_{ij}
+2\pi v_{ij})^2}\\
&=\sum_{v_{ij}=-\infty}^{\infty}\sqrt{\tau_0\over{2\pi E_J}}
\int dp_{ij}e^{-{{p_{ij}^2\tau_0}\over{2E_J}}+
i{ p}_{ij}\tau_0 (\phi_{ij}-\vec A_{ij}+2\pi v_{ij})}
\end{array}
\label{villain}
\end{equation}
This approximation is valid for  $E_J\tau_0\widetilde{>}1$, and gets better as
$E_J\tau_0$ gets larger. However, it retains the most important
feature of the Josephson energy, the periodicity with respect to
$\phi$, for all values of $E_J\tau_0$.
Alogether, then, the approximation we
discuss holds  for $E_J>E_C$. The significance of the
 field $\vec v_i$ can be understood
by noting that
${\vec\Delta}\times{\vec v}_i$  describes the density of
vortices \cite{Polyakov}.
As for
the real variable $ p_{ij}$, as shown below, it
describes the Josephson current along the bond $ij$.
Since the Josephson energy includes a sum over all lattice bonds, the
Villain approximation introduces a variable $p_{ij}$ to each lattice bond.
Thus, we can regard $p$ as a {\it vector}
defined for each lattice $site$, such that $p_{ix}$ corresponds to
the bond $i,x$ and $p_{iy}$ corresponds to the bond $i,y$.
Similarly, the difference $\phi_i-\phi_j$,
 the integral $\int_i^j{\vec A}\cdot dl$ and the variables $v_{ij}$
can be regarded as
vectors ${\vec\Delta}\phi_i$, ${\vec A}_i$ and $\vec v_i$.

Next we apply the Poisson resummation formula to the $n_i$ dependent part
of the action. By doing that we make the $n_i$ variables real numbers
rather than
integers and add a time component to the integer--valued
vector field ${\vec v}_i$.
The partition function then becomes,
\begin{equation}
\begin{array}{lll}
Z=\sum_{\{{\bf{ v}_i(t)}\}}
&\int  D\{n_i(t)\}\int D\{{\vec p}_i(t)\}\int \{D\phi_i(t)\}&\\ \\
&\exp\Bigg\{\int_0^\beta dt\Big[
\sum_iin_i({\dot\phi}_i+ 2\pi
v_{0i})&-{{(2e)^2}\over 2}{\sum_{ ij}} (n_i-n_x){\hat C}^{-1}_{ij}(n_j-n_x)
\\ \\ & &-\sum_i{{{\vec p}_i^2}\over{2 E_J}}+
i{\vec p}_i\cdot({\vec\Delta}\phi_i-{\vec A}_i+2\pi{\vec v}_i)\Big]\Bigg\}
\end{array}
\label{poisson}
\end{equation}
where  the path integral should be performed stepwize \cite{Swanson}.
For the brevity of this expression we omitted the explicit time dependence
of $n_i,\phi_i,{\vec p}_i, {\vec v}_i$ in the stepwize
integrated action.
This form allows us to understand the physical significance of $\vec p$.
The only $\vec A$--dependent term in the action is $-i{\vec p}_i\cdot
{\vec A}_i$. The derivative of the Lagrangian with respect to $\vec A$ is
the current. Thus, $\vec p_i$ is the Josephson current flowing through the
site $i$.

The path integral over the phase variables $\phi_i(t)$
 can now be performed. The phase
$\phi_i$ at the site $i$ is coupled to the charge $n_i$ (via the term
$in_i{\dot\phi_i}$) and to the vector $\vec p$ at the site $i$ and
its nearest neighbors.
 The integration over $\phi_i$ yields conservation of charge
 constraint on the integration
 over $n_i,{\vec p}_i$, in the form
\begin{equation}
\Delta_tn_i+{\vec\Delta}\cdot{\vec p}_i=0
\label{dcoc}
\end{equation}
where the definition $\Delta_tn_i\equiv
{1\over\tau_0}[n_i(t+\tau_0)-n_i(t)]$
makes the difference operator $\bf\Delta$ a 3--vector.

 The constraint (\ref{dcoc})
 is nothing but a discretized form of a two dimensional
 conservation of charge equation.  Like the latter,
 it can be solved by defining a 3--vector
field $\bf{\cal K}$ that satisfies,
\begin{equation}
\epsilon^{\alpha\beta\mu}\Delta_\alpha{{\cal K}}_{i,\beta}=2\pi{\bf p}_{i,\mu}
\label{curl}
\end{equation}
 where ${\bf p}_i
\equiv(n_i, p_{i,x}, p_{i,y})$.   The three
components of $\bf{\cal K}_i$ are real, like those of
${\bf p}_i$. The definition (\ref{curl}) of $ {\cal K}_i$ is
not unique and it becomes unique only when a gauge is fixed. The partition
function is, of course, independent of that gauge.
The constrained  path integral  over
${\bf p}_i$ is replaced now by  path integrals
 over $\bf{\cal K}_i$, constrained by the gauge condition.
Here we choose to work in the Coulomb gauge, in which
${\vec\Delta}\cdot{\vec{\cal K}}=0$. In that gauge  the partition function
becomes,
\begin{equation}
\begin{array}{ll}
Z=\sum_{\bf v_i(t)} \int  D\{{\bf{\cal K}_i}\}
&\exp{\int_0^\beta dt}  \left[-{{e^2}\over {2\pi^2}}
{\sum_{ ij}} ({\vec\Delta}\times{\vec{\cal K}}_i)
{\hat C}^{-1}_{ij}
({\vec\Delta}\times{\vec{\cal K}}_j)\right.
\\ \\  & \left. -\sum_i{1\over{8\pi^2 E_J}}(\Delta_t{\vec{\cal K}}_i^2
+{\vec\Delta}{\cal K}_{0i}^2)+i\sum_i[{\bf\Delta}\times({\bf{\cal K}}+
{\bf{\cal K}}^{ext})]_i\cdot({\bf v_i}-{1\over{2\pi}}{\bf A}_i)
\right]
\end{array}
\label{vilsav}
\end{equation}
where ${\vec{\cal K}}^{ext}$ is defined by ${\vec\Delta}
\times{\vec{\cal K}}^{ext}=2\pi
n_x$.
We are now one step away from having an effective action for the
vortices. The remaining step is an integration by parts of the last
term in the action in (\ref{vilsav}). After performing that integration,
 the following action is obtained:
\begin{equation}
\begin{array}{ll}
S^{vor}=\int_0^\beta dt
\sum_i\Big\{ &i(\rho^{vor}_i-{\bar n_v}){\cal K}_{0i} +i{\vec J}^{vor}_i
\cdot({\vec{\cal K}}_i+{\vec{\cal K}}^{ext}) \\ \\
&+  {{e^2}\over{2\pi^2\hbar^2}}\sum_j ({\vec\Delta}\times{\vec{\cal K}}_i)
{\hat C}^{-1}_{ij}
({\vec\Delta}\times{\vec{\cal K}}_j)+{1\over{8\pi^2 E_J}}
((\Delta_t{\vec{\cal K}}_i)^2
+({\vec\Delta}{\cal K}_{0i})^2)\Big\}
\end{array}
\label{actiona}
\end {equation}
where the vortex 3--vector current ${\bf J^{vor}}$ is defined as
${\bf J^{vor}}={\bf\Delta}\times{\bf v}$, the average density of
vortices is given by ${\bar n}_v={B\over\Phi_0}$ and the value of
$\hbar$ has been restored.
Equation (\ref{actiona}) is the
starting point of the discussion in section (3).

\vskip 1cm \noindent
{\large\bf  Acknowledgements}

I am indebted to B.I. Halperin, S. Simon and  D.H. Lee for instructive
discussions, and to M.Y. Choi for sending me his preprint prior to
publication. I am grateful to the Harvard Society of Fellows for
financial support. Part of this work was done in the Aspen center of
physics, to which I am grateful for hospitality. Part of this work was
supported by NSF Grant No. DMR-91-15491.
\\ \\


\begin{thebibliography}{99}
\bibitem{Notea} The magnitude of the gap in the excitation spectrum of
$\vec{\cal K}$ depends on the precise form of $\epsilon_J({\vec r})$.
However, its
existence does not.
\end{thebibliography}
\end{document}